\documentstyle[epsfig, twocolumn]{iso98}

\newcommand{\thepapername}{}

\begin{document}

\setlength{\parindent}{0pt}
\setlength{\parskip}{ 10pt plus 1pt minus 1pt}
\setlength{\hoffset}{-1.5truecm}
\setlength{\textwidth}{ 17.1truecm }
\setlength{\columnsep}{1truecm }
\setlength{\columnseprule}{0pt}
\setlength{\headheight}{12pt}
\setlength{\headsep}{20pt}
\pagestyle{veniceheadings}

\title{\bf MEDIUM-DISTANT CLUSTERS OF GALAXIES SEEN BY ISO\thanks{ISO is an ESA
project with instruments funded by ESA Member States (especially the PI
countries: France, Germany, the Netherlands and the United Kingdom) and
with the participation of ISAS and NASA.}}

\author{{\bf L.~Lémonon, M.~Pierre, C.J.~Cesarsky, D. Elbaz, L. Vigroux} \vspace{2mm} \\
DAPNIA/SAp, CEA Saclay, France 
}

\maketitle

\begin{abstract}

We present a comprehensive study of ISO cluster galaxies in connection with other wavelengths. First results show that infrared galaxies detected by ISOCAM concentrate on the edge of clusters. Surprisingly, they are not emission lines galaxies but very luminous elliptical, merging or interacting ones, which suggests non-trivial connections between star formation rate, optical and mid-IR properties. However, no correlation has been found between the MIR/optical star formation properties of the galaxies, so that the exact origin of the IR emission remains unclear.
\vspace {5pt} \\

  Key~words: ISO; infrared astronomy; galaxy clusters.

\end{abstract}

\section{INTRODUCTION}

Until recently, detailed cluster studies were restricted to nearby clusters ($z < 0.1$). Nowadays, new imaging as well as spectroscopic capabilities (visible, X-ray, IR, radio) have considerably improved our view of the distant universe. 
In this context, we have selected a sample of ten bright objects, expected to be rich clusters, from the {\em ROSAT All Sky Survey} (RASS) in the $z = 0.15 - 0.3$ range in order to investigate issues ranging from cluster formation and evolution down to a local phenomenon such as the enrichment of the ICM (\cite{Pierre94}). We have undertaken a detailed high-resolution multi-wavelength follow-up including ROSAT HRI and PSPC, ASCA pointing-s, radio images at 3, 6, 13, 20 and 36 cm and extensive photometry and spectroscopy in the optical. Seven of them were included in the ISO Central Program (DEEPXSRC, 30 hours, P.I.: C.J.\ Cesarsky; Collaborator: M.\ Pierre), in order to study the influence of the environment on the galaxy evolution.
We present here the analysis which allowed us to statistically determine properties of IR emitting galaxies in our clusters.

\section{OBSERVATIONS}

 Seven objects extracted from the RASS have been observed by ISO, at 6.75 and 15\,$\mu$m with ISOCAM and at 90\,$\mu$m with ISOPHOT. From our X-ray/radio/optical, four clusters turned out to be very rich objects ranging from $z = 0.165$ (Abell 1111) to $z = 0.307$ (Abell 1300), and three objects appeared to be small groups with a central AGN (see Table \ref{isoobs}) (\cite{these}, hereafter L99). All observations were made with a pixel size of 6 arcsec, and each sky position as observed four times. The various environmental conditions are represented in those seven objects, fully described in several papers, and can be summarized as follow :
\begin{itemize}
\item Abell 1300 : a post merging cluster containing a radio halo and a relic
radio source (\cite{Lemonon97}, \cite{Reid98})
\item RXJ1104-2330 : a strong AGN of BL-Lac type possibly embedded in a small cluster or a large group of galaxies (L99)
\item Abell 1732 and Abell 1111 : two cooling flow clusters respectively large and small (\cite{}, L99)
\item RXJ1032-2607 : probably a strong AGN in a small group of galaxies (L99)
\item RXJ1314-2516 : a new very interesting merging cluster of galaxies, containing probably a relic radio source (L99)
\item RXJ1315-4236 : a more nearby ($z = 0.105$) AGN of BL-Lac type, in a group of galaxies (L99)
\end{itemize}

\begin{table}[htb]
  \caption{\em DEEPXSRC ISOCAM observations.
The column are :
(1) the object name from the ACO catalogue (Abell et al. 1989) when existing or the ROSAT All Sky Survey number. 
(2) the redshift of the object. 
(3) the size of the side of the maximum sensitivity box area. 
(4) the sensitivity in the LW2 ($6.75 \mu$m) CAM filter. 
(5) the sensitivity in the LW3 (15 $\mu$m) CAM filter (RXJ1314-2516 was not
observed in the LW3 band).}
  \label{isoobs}
  \begin{center}
    \leavevmode
    \footnotesize
    \begin{tabular}[h]{ccccccccc}
      \hline \\[-5pt]
      Object & z      & Field  & LW2  & LW3 \\[+5pt]
      name & & of View & $\mu$Jy & $\mu$Jy\\
      \hline \\[-5pt]
      Abell 1300   & 0.307 & $6.4'$ & $500$ & $1400$\\
      RXJ1104-2330 & 0.187 & $6.4'$ & $450$ & $1000$\\
      Abell 1732   & 0.193 & $6.4'$ & $400$ & $1000$\\
      Abell 1111   & 0.165 & $6.4'$ & $500$ & $1300$\\
      RXJ1032-2607 & 0.247 & $4.8'$ & $750$ & $1500$\\
      RXJ1314-2516 & 0.244 & $4.8'$ & $500$ & - \\
      RXJ1315-4236 & 0.105 & $9.6'$ & $750$ & $1100$\\
      \hline \\
      \end{tabular}
  \end{center}
\end{table}

\section{IR EMISSION IN ISOCAM FILTERS}

As discussed by \cite*{Fadda99}, the k-correction for distant clusters is
a very important point when we look in the mid-IR wavelength. In our redshift
range ($0.165 < z < 0.307$), LW2 always includes only the $6.2 \,\mu$m Unidentified Infrared Band (UIB) feature, but is more and more (with the redshift) contaminated by the rayleigh-jeans tail of red stars. The LW3 filter is less and less sensitive to the warm dust but is always contaminated by the $11.3 \mu$m UIB feature and the $9.7 \mu$m silicate absorption. Unfortunately, the interpretation of mid-IR properties is turning into a maze by this mixity of IR contributions.

\section{RESULTS}

\subsection{What Kind Of Objects Did We Detect ?}

First, two third of the objects detected by ISOCAM are optically identified as  stars. On about 70 remaining galaxies, since we had optical spectra for 35 of them, we were able to evaluate the contamination by field galaxies. One third of the 35 appeared not to belong to the clusters, so by extrapolation, we expect that some 50 out of 70 are really clusters members. At our depth,
20 fields galaxies on the total area covered by our ISOCAM observation are in a good agreement with deep $15 \,\mu$m count at 1 mJy from ELAIS (\cite{Rowan}). We detect as many galaxies at 6.75 $\mu$m as at 15 $\mu$m although very few of them are detected in both filters (about 15\%). Detected galaxies are bluer (V-I $\sim 1.2$) than the mean cluster galaxy (V-I $\sim 1.5$, which corresponds to "normal" early-type galaxies at those redshifts) but their spectrum look like elliptical galaxies.

\subsection{Radial Distribution Of ISOCAM Galaxies In Clusters}

\begin{figure}[ht]
  \centerline{\epsfig{file=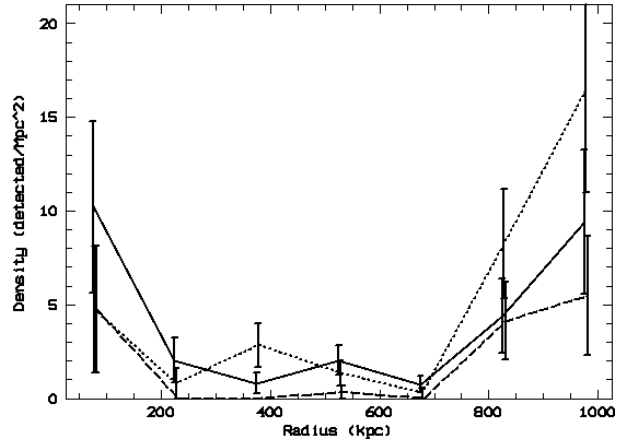,width=8cm}}
  \caption{\em Radial distribution of ISOCAM galaxies in a composite cluster
  made of all detected galaxies in each cluster. Solid and dotted lines show $6.75$ and $15 \mu$ galaxies, respectively, dashed line shows galaxies detected in both bands.}
  \label{distriso}
\end{figure}

Figure \ref{distriso} shows the radial distribution of a composite cluster made of all detected galaxies. The density in the center is only due to the central dominant (cD) galaxies which are 10-100 times more luminous in the optical that normal elliptical galaxies and thus expected to be more luminous also in the mid-IR. The main observation is that most of detected galaxies concentrate on the edge of the cluster. Below 750 kpc, only one galaxy is detected in both band (cD galaxies excluded) although about 20 galaxies are detected at 6.75 $\mu$m or $15 \mu$m. This may suggests that 6.75 $\mu$m and 15 $\mu$m galaxies are two different populations of galaxies. To the contrary, beyond 750 kpc, more than 50\% of them are detected in both bands, so galaxies in this range may be
a different population. The seven galaxies detected in both bands are bluer than the mean detected galaxies with an optical color of V-I $\sim 0.95$, corresponding to Scd type.

\subsection{Galaxies Detected Only At 6.75\,$\mu$m}
\label{lw2}

\begin{figure}[ht]
\centerline{\psfig{file=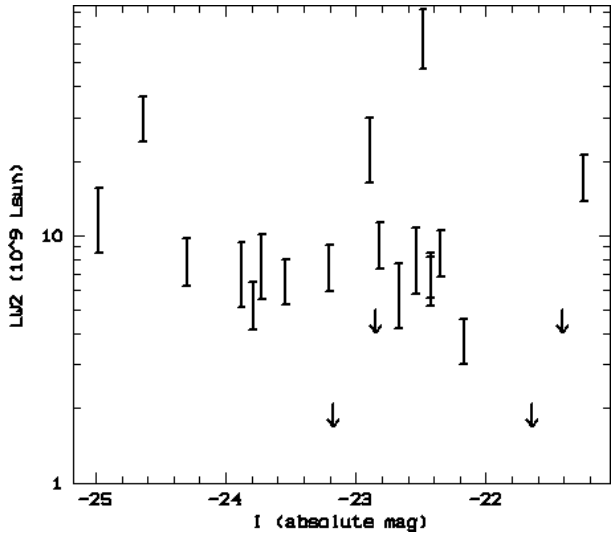,width=8cm}}
\caption[]{The 6.75 $\mu$m ($\mu$Jy) luminosity in function of the I absolute magnitude for objects detected only at 6.75 $\mu$m. Four galaxies are well outside the
general trend, they must be field galaxies or dusty/star-forming objects.}
\label{lw2magI}
\end{figure}

This population may be the signature of the rayleigh-jeans tail of an old stars population, as :
\begin{enumerate}
\item all cD galaxies are detected, and they are the most luminous galaxies of
each clusters,
\item there is likely a correlation between LW2 and I flux (see Figure \ref{lw2magI}), 
\item the mean V-I color ($\sim 1.4$) is closer of "normal" elliptical galaxy
than others detected objects,
\item regarding our 15 $\mu$m completeness, we can be sure that most of galaxies detected at 6.75 $\mu$m have a 15 $\mu$m flux {\em below} the 6.75 $\mu$m one, excluding intense star formation activity.
\end{enumerate}
Even if many galaxies seen only at 6.75 $\mu$m are detected beyond 750 kpc where elliptical galaxy density is much lower than in center, the low total number of them is fully compatible with the statistically normal density of early-type in outskirt regions.

\subsection{Galaxies Detected At 15\,$\mu$m}

It is well known that spiral galaxies in clusters are more concentrated in the external regions (\cite{Dressler80}). Figure \ref{distriso} shows that galaxies detected at 15 $\mu$m are preferentially seen beyond 750 kpc, and much more than 6.75 $\mu$m ones. Half of them are detected also at 6.75 $\mu$m with flux ratio typical of photo-dissociation regions or star formation activity. Their optical colors are bluer than mean cluster galaxies.

\subsection{Effects Of The Environmental Conditions}

It could be a correlation between the number of detect galaxies at $6.75 \,\mu$m and the X-ray luminosity in the [0.1-2.4 keV] band (L99). Since the X-ray
luminosity is related to the richness of the clusters, this corroborates the
suggestion of section \ref{lw2} as most LW2 galaxies are luminous early-type galaxies. Other effects of environmental conditions are very hard to point out because of our low statistic (between 4 and 15 galaxies detected by ISOCAM per cluster).

\subsection{Optical Correlations With Star-Forming Galaxies}

\begin{figure*}[ht]
\centerline{\psfig{file=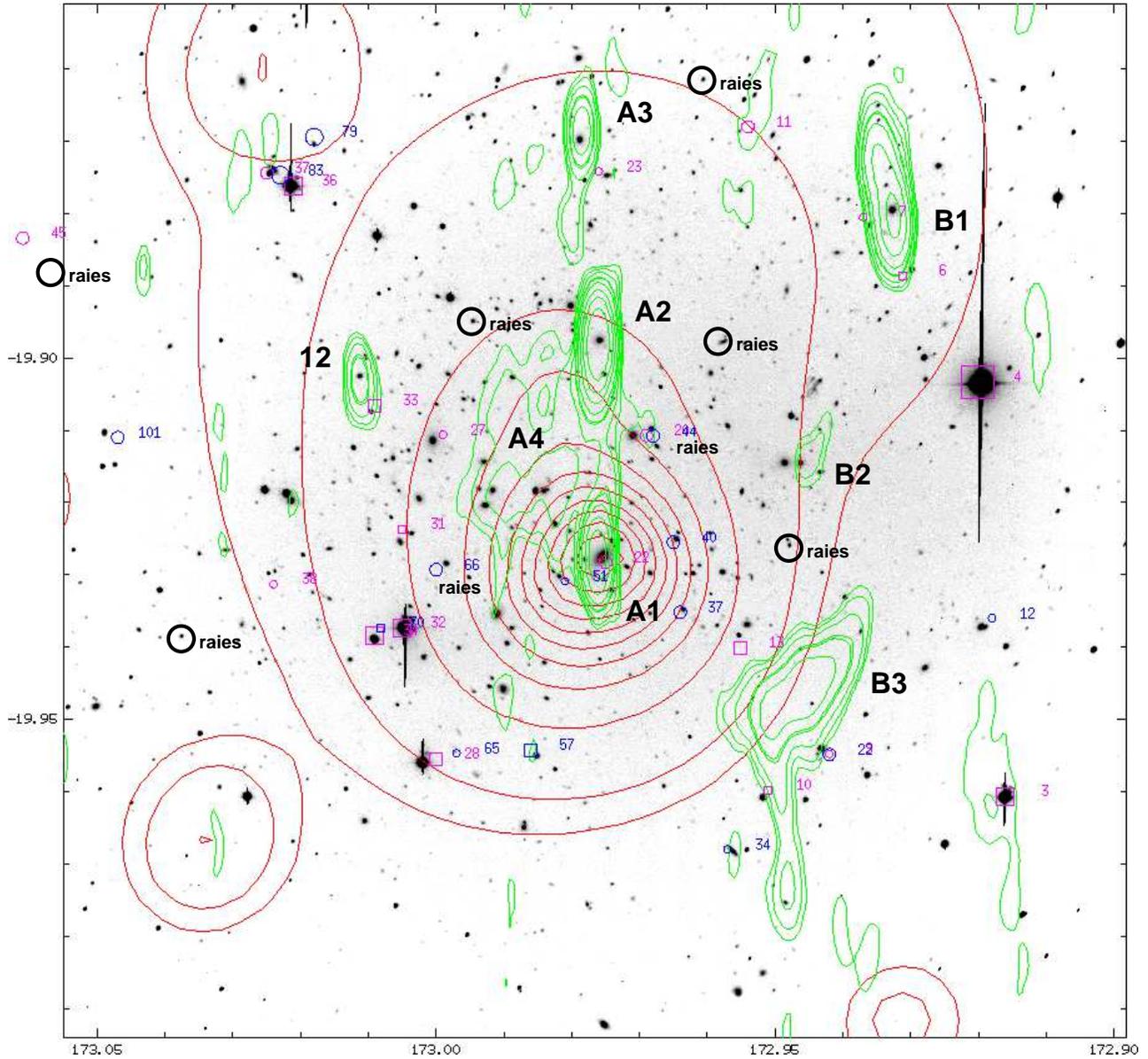,width=17cm}}
\caption[]{Abell 1300, a post-merging cluster (\cite{Lemonon97}, \cite{Reid98}). On the R band image taken on the CFHT are overlaid in
red contours X-ray [0.4-2.4 KeV] PSPC levels ($8.3 \times 10^{-4}$, and $1.8 \times 10^{-3}$ to $6 \times 10^{-2}$ by step of $5.76 \times 10^{-3}$ counts/s/arcmin$^2$)and in green radio (1.4 GHz) ATCA levels ($0.2, 0.3, 0.5, 0.9, 1.6, 3.2$ and $6$ mJy/beam). It confirms that detected ISOCAM galaxies (in magenta and blue for respectively $6.75 \,\mu$m and $15 \,\mu$m bands, circle for galaxies and squares for point-like objects, with size proportional to emitting fluxes) in clusters above a limited flux are not galaxies with emission lines but very luminous elliptical, merging or interacting galaxies (\cite{Pierre96}). Line-emitting galaxies are pointed by the "RAIES" label. Abell 1300 contains the most distant radio halo known (A4) and a relic radio source (B3). This is one out of our seven objects, others are available in L99.}
\label{overlays}
\end{figure*}

In the course of our optical follow-up, we have discovered that detected ISOCAM galaxies in medium distant clusters (at a sensitivity of $\sim$ 1 mJy) are not emission lines galaxies but very luminous elliptical, merging or interacting ones, which suggests non-trivial connections between star formation rate, optical and mid-IR properties (see Figure \ref{overlays}). However, the spectral resolution at which the redshift of these galaxies were measured (15 \AA) does not allow us to identify any peculiar signature in the optical spectra of these objects, so that the exact origin of the IR emission remains unclear.
This observation became a very scheming subject since IR deep surveys claim the existence of a population of galaxies with great IR fluxes in the past ($0.6 < z < 1$) with no peculiar optical signature (\cite{Elbaz98}, \cite{Aussel98}) and could be a new starting point to the comprehension of the stars formation history. 

\section*{ACKNOWLEDGMENTS}
The ISOCAM data presented in this paper was analysed using "CIA", a joint development by the ESA Astrophysics Division and the ISOCAM Consortium. The ISOCAM Consortium is led by the ISOCAM PI, C. Cesarsky,  Direction des Sciences de la Matiere, C.E.A., France.


\begin{thebibliography}{}
\bibitem[\protect\astroncite{Abell et~al.}{1989}]{ACO}
Abell, G.O., Corwin, H.G., Olowin, R.P., 1989, ApJS {\bf 70}, 1
\bibitem[\protect\astroncite{Aussel et~al.}{1998}]{Aussel98}
Aussel, H., Cesarsky, C., Elbaz, D., Starck, J.L., 1998, A\&A in press, astro-ph/9810044
\bibitem[\protect\astroncite{Dressler}{1980}]{Dressler80}
Dressler, A., 1980, ApJ {\bf 326}, 351
\bibitem[\protect\astroncite{Elbaz et~al.}{1998}]{Elbaz98}
Elbaz, D. et al., 1998, "NGST" 34th Liege Astrophysics Colloquium, eds.\, Benvenuti P. et al., ESA Pub., astro-ph/9807209
\bibitem[\protect\astroncite{Fadda \& Elbaz}{1999}]{Fadda99}
Fadda, D. \& Elbaz, D., 1999, this book
\bibitem[\protect\astroncite{Lémonon et~al.}{1997}]{Lemonon97}
Lémonon, L., Pierre, M., Hunstead, H., Reid, A., Mellier, Y., B\"ohringer, H., 1997, A\&A {\bf 326}, 34
\bibitem[\protect\astroncite{Lémonon et~al.}{1999}]{Lemonon99}
Lémonon, L., Pierre, M., Cesarsky, C.J. et al., 1999, in preparation
\bibitem[\protect\astroncite{Lémonon}{1999}]{these}
Lémonon, L., 1999, PhD. thesis, University of Paris XI, France
\bibitem[\protect\astroncite{Pierre et~al.}{1994}]{Pierre94}
Pierre, M., B\"ohringer, H., Ebeling, H., Voges, W., Schuecker, P. et al., 1994, A\&A {\bf 290}, 725
\bibitem[\protect\astroncite{Pierre et~al.}{1996}]{Pierre96}
Pierre, M., Aussel, H., Altieri, B. et al., 1996, A\&AL {\bf 315}, L297
\bibitem[\protect\astroncite{Reid et~al.}{1998}]{Reid98}
Reid, A.D., Hunstead, R.W., Lémonon, L., Pierre, M., 1998, MNRAS accepted
\bibitem[\protect\astroncite{Rowan-robinson et~al.}{1999}]{Rowan}
Rowan-robinson, M. et al., 1999, this book
\end{thebibliography}
\end{document}